\documentclass[]{article}

\usepackage{amssymb,latexsym,amsmath}     
\usepackage{graphicx}      
\usepackage{multirow}   
\usepackage{array}
\usepackage{enumitem}
\usepackage{pgfplots}
\usepackage{tikz}
\pgfplotsset{width=7cm}
\renewcommand{\arraystretch}{1.5}

\usepackage[english]{babel}
\usepackage[affil-it]{authblk}

\begin{document}
\bibliographystyle{}
\bibliography{bibli}
\title{Risk aversion and Catastrophic risks: the pill experiment}
\author{Julien BLASCO
	\thanks{Electronic address: \texttt{julien.blasco@ensta-paristech.fr}}}
	\affil{\'Ecole Nationale Sup\'erieure de Techniques Avanc\'ees \footnote{ENSTA ParisTech: http://www.ensta.fr/}
	 \\ Columbia Consortium for Risk Management,
	 \\ Columbia University, NY, USA (summer research)}
	
\author{Graciela CHICHILNISKY
	\thanks{Electronic address: \texttt{chichilnisky1@gmail.com}}}
	\affil{Columbia Consortium for Risk Management,
	 \\ Columbia University, NY, USA}
\maketitle

\begin{abstract}
This article focuses on the work of O. Chanel and G. Chichilnisky (2013) \cite{chanel} on the flaws of expected utility theory while assessing the value of life. Expected utility is a fundamental tool in decision theory. However, it does not fit with the experimental results when it comes to catastrophic outcomes ---see, for example, Chichilnisky (2009) \cite{chichifear} for more details. In the experiments conducted by Olivier Chanel in 1998 and 2009, several subjects are ask to imagine they are presented 1 billion identical pills. They are paid \$220,000 to take and swallow one, knowing that one out of 1 billion is deadly. The objective of this article is to show that risk aversion phenomenon cannot explain the experimental results found. This is an additional reason why a new kind of utility function is necessary: the axioms proposed by Graciela Chichilnisky will be briefly presented, and it will be shown that it better fits with experiments than any risk aversion utility function.
\end{abstract}


\newpage

\section{Introduction}
One of the most commonly used frameworks for describing choices by economic agents is Expected Utility. But what if one of the outcomes is death? Can expected utility still be consistent with real-life choices? This question is fundamental when it comes to decision implying catastrophic outcomes. Based on the experiments conducted by Chanel and Chichilnisky (2013) \cite{chanel}, issues raised by the use of expected utility when confronted to rare and catastrophic events are discussed.  \\

In \cite{chanel}, several subjects are presented 1 billion identical pills. They are paid \$220,000 to take and swallow one, knowing that one out of 1 billion is deadly. In this experiment conducted in 1998, about half of the subjects accept the deal, and the other half refuses it. The following symbols will be used: the subject has an initial wealth $l$, which corresponds to their life and of which the value is yet to be determined. By accepting the deal, the subject is rewarded with an amount $r=220000$. With a probability $p=10^{-9}$, the subject dies (their wealth becomes $r$), and with a probability $1-p$, the subject survives (hence their wealth becomes $l+r$).\\

\section{Using mean gain: linear preferences}
\label{mean}
At first, in \cite{chanel}, a naive approach is adopted. It is assumed that a subject who refuses the deal automatically values their life at more than $l_0=220000/10^{-9}$. Why? Because the subject thinks that the value of their life is greater than the mean gain $m(p)$ they will get by taking the deal. This can be expressed as follows:
\begin{gather*} m(p)=pr+(1-p)(l+r)<l \\
			 \text{i.e.}\quad r<pl \\
			 \text{which is equivalent to}
\end{gather*}
\begin{equation}\label{l0}l>\frac{r}{p}=l_0 \end{equation}

The value of $l_0$ is much greater than the value of life that is generally admitted in literature, i.e.\ between \$1.7 and \$7 million \cite{chanel}. The singularity of the experiment results will be explained later on.
\\

The computation which leads to \eqref{l0} implicitly makes the assumption that the utility function $u$ is linear. Indeed, in expected utility theory, an economic agent makes choices regarding the utilities of every outcome, not their gross values. Hence, the only equation that can be inferred from the fact that the subject refuses the deal is what follows:
\begin{equation}pu(r)+(1-p)u(l+r)<u(l)\end{equation}
However, the result \eqref{l0} is still valid if the utility function is linear. But this is not generally true. By assuming that hypothesis, one of the effects that could explain the singular experimental results is eliminated. Indeed, risk aversion phenomenon could be the reason why subjects do not want to take a chance by choosing a pill, even if the mean gain is greater than the value they assign to their own lives.\\

Let us focus on a subject who would not mind accepting the deal, that is to say who gets the same satisfaction either while swallowing the pill, or while refusing the deal. The following equation is implied by this statement:
\begin{equation} \label{debut} pu(r)+(1-p)u(l+r)=u(l) \end{equation}

\section{Utility functions with risk aversion}

Risk aversion can simply be summed up by the famous saying ``a bird in hand is worth two in the bush''. An economic agent will generally prefer getting \$100 in hard cash rather than playing head or tails and have 1 chance out of 2 to gain \$200. In general, if the mean value of a lottery $p$ is $v$, then the utility of $v$ will be greater than the expected utility of the lottery $p$:
\begin{equation}\label{jensen}u(v)>EU(p)\end{equation}
In terms of utility function, this is equivalent to say that $u$ is a concave. It is thus clear that inequation \eqref{jensen} is simply a Jensen inequality. Risk aversion can be assessed, thanks to Arrow-Pratt Absolute Risk Aversion (ARA) measure $A$, which is defined as follows:
\begin{equation}A= -\frac{u''}{u'} \end{equation}
where $u'$ is the first derivative of $u$, and $u''$ its second derivative. If the preferences are increasing and concave, $A$ is positive.

\subsection{Search of a consistent utility function}
The objective of this section is to find a utility function which explains the results of the pill experiment. Some famous models will be proposed and their parameters adjusted. Thanks to equation \eqref{debut}, for every utility function that is used, the implied value of life $l$ can be found. The tests will be made with two kinds of utility function. The results are summed up in the following table \ref{life}, which indicates the values of the parameter $\gamma$ for which $l$ is within the borders of the value of a statistical life (i.e.\ between \$1.7 and \$7 million). \\

\begin{table}[h]
\caption{Value of life depending on the utility chosen}
\begin{tabular}{r|c|c|c|}
\cline{2-4}
&\multicolumn{2}{ c| }{expressions of u} & value of life  \\ \cline{2-4}
linear preferences& \multicolumn{2}{ c| }{$u(x)=ax+b$}& $220000/10^{-9}$ \\ 
\cline{2-4}
\multirow{2}{*}{constant ARA}&$u(x)=-e ^{-\gamma x}$&  $\gamma =10^{-5.53}$  &  $l=7.0\times 10^{6}$  \\
\cline{3-4}
&$A=\gamma$  & $\gamma =10^{-4.86}$ & $l=1.7\times 10^{6}$ \\ 
\cline{2-4}
\multirow{2}{*}{variable ARA}& $u(x)=-x^{-\gamma}$ & $\gamma =5.3 $& $l=7.0\times 10^{6}$ \\
\cline{3-4}
& $A=(1+\gamma)/x$ & $\gamma =10 $ & $l=1.7\times 10^{6}$ \\
 \cline{2-4}
\multicolumn{4}{c}{ }
\end{tabular}
\label{life}
\end{table}

These are two examples of functions that can explain why some people refuse the deal. They do not necessarily value their lives at more than \$220 trillion: maybe they just have a high risk aversion. It would be interesting to see whether those utility functions are consistent with other lotteries.

\subsection{Test of various utility functions}

Let us build a very simple test: this is a classic ``head or tails'' lottery. Head, the subject wins \$100. Tails, he wins \$200. This is an interesting deal indeed. The question is: how much is a subject ready to pay for such a lottery? The following table shows the results, for the three reference functions.

\begin{table}[h]
\centering
\caption{Value of the head or tails deal}
\begin{tabular}{|c|c|}
\hline
expressions of $u$ & value of the deal \\
\hline
$u(x)=ax+b$ & 150 \\
\hline
$u(x)=-e^{-x\times 10^{-5}}$ & 149.98 \\
\hline
$u(x)=-x^{-7}$ & 110.3 \\
\hline
\end{tabular}
\end{table}

The third function seems inconsistent with day-to-day experiments: its risk aversion is too high. A random subject would generally prefer to play the game rather than being paid \$110.3. \\

The exponential function, though, gives a relatively good result with respect to the mean gain of the deal. For now, this seems to be a consistent utility function which could explain the results of O. Chanel and G. Chichilnisky's experiments on the value of life.

\section{The non-takers}

\subsection{Failure of expected utility}

During the 1998 experiment and the one that followed in 2009, some subjects not only refused the deal, but would not take a pill no matter what, and regardless of the probability of the deadly pill. This is a response that cannot be explained by standard expected utility.   \\ 

The utility function is necessarily increasing, so $u(l+r)>u(l)$. Hence:

\begin{equation}
\label{limit}
\lim_{p\rightarrow 0}EU(p)=\lim_{p\rightarrow 0}\left[ pu(r)+(1-p)u(l+r)\right] =u(l+r)>u(l) \end{equation}

The conclusion is that for every kind of utility function, and regardless of risk aversion, there will always be a small enough probability $p$ --or a big enough stack of pills-- such that the subject will accept to take the pill. The table \ref{table:pill} shows the probability thresholds for different risk aversions, given a value of life of \$2 million (the deal is acceptable if its value is over \$2 million).
\begin{table}[h]
\centering
\caption{Value of the pill deal}
\label{table:pill}
\begin{tabular}{|c|c|c|}
\hline
$u(x)=-e^{-\gamma x}$ where $\gamma =$... & probability $p$ & value of the deal \\
\hline
$10^{-5}$ & $10^{-9}$ & $2.18\times 10^6$ \\
\hline
$10^{-4.9}$& $10^{-10}$& $2.04\times 10^6$ \\
\hline
$10^{-4.8}$& $10^{-13}$ & $2.10\times 10^6$\\
\hline
\end{tabular}
\end{table} 
\\

The following graph describes the evolution of the value $v$ given to the pill deal according to the probability $p$ of finding the deadly pill.\\

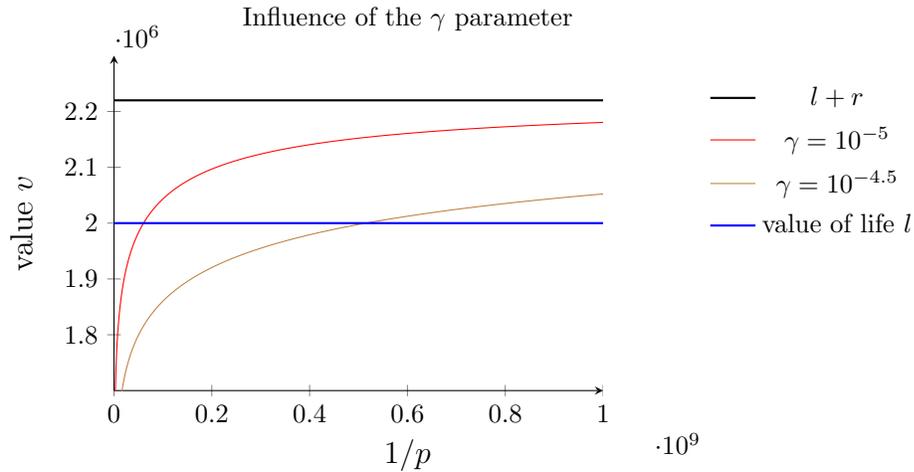
\begin{tikzpicture}
\begin{axis}[xscale=1.2,
	axis x line=bottom,
	xlabel={\large{$1/p$}},
	ylabel={\large{value $v$}},
	title={Influence of the $\gamma$ parameter},
	axis y line = left,
	xmin=0,xmax=1e9,
	ymin=1.7e6,ymax=2.3e6,
	ytick={1.8e6,1.9e6,2.0e6,2.1e6,2.2e6},
	legend style={at={(1,0.94)},anchor=north west,draw=none,row sep=0.1cm},
	legend entries={$l+r$,$\gamma = 10^{-5}$,$\gamma=10^{-4.5}$,value of life $l$}
]

\addplot[
domain=0:1e9,
thick
]
{ 2220000};

\addplot[
red,
domain=0:1e9,
samples=500,
]
{ -1e5 * ln (-( -1/x * exp(-220000*1e-5) - (1-1/x)*exp(-1e-5 * 2220000) )};

\addplot[
brown,
domain=0:1e9,
samples=500,
]
{ -1/10^(-4.95) * ln (-( -1/x * exp(-220000*10^(-4.95)) - (1-1/x)*exp(-10^(-4.95) * 2220000) )};

\addplot[
blue,
thick,
domain=0:1e9,
]
{ 2000000};

\end{axis}
\end{tikzpicture}

The above data is not consistent with the results obtained with part of the subjects. Indeed, some people will not accept to play under any conditions, due to the catastrophic outcome that is death. Classic expected utility tends to under-estimate the impact of such catastrophic events on our decision-making (see \cite{chanelfear} for other experimental evidence). \\

There is a need for a utility function that takes into account catastrophic events, even when probabilities are very low.

\subsection{Taking into account catastrophic events}

\subsubsection{A new kind of utility function}

In Chichilnisky (2009) \cite{chichifear}, a new kind of utility function is proposed, which takes into account events with arbitrarily small probabilities, but with huge effects. To understand precisely what modifications has to be done to the classical expected utility, one has to define explicitly the underlying mathematical framework. \\

A lottery is an essentially bounded function $f\in L_\infty (\mathbb{R})$ which represents the utility of every outcome $x\in \mathbb{R}$. In order to rank the different lotteries, expected utility is defined by $W(f)=\int_\mathbb{R} f(x)d\mu(x)$ where $\mu$ is a measure with an integrable density function $\phi_1\in L_1(\mathbb{R})$. $W$ is a linear functional on the lottery space. $f$ is said to be ``preferable to $g$'' if and only if $W(f)>W(g)$. The issue is that this ranking is insensitive to rare events: it means that if a catastrophic outcome is added to the lottery $f$, and its probability is low enough, the ranking will not be modified. \\

Before introducing the new utility function found by G. Chichlnisky, one has to remember that any linear functional on $L_\infty(\mathbb{R})$ can be expressed as a Radon integral, with respect to a measure that is absolutely continuous with respect to Lebesgue measure \cite{french}. However, those Radon integrals can be expressed as Lebesgue integrals if and only if that measure is $\sigma$-additive. Otherwise (i.e.\ if the measure is additive but not $\sigma$-additive), the result cannot be expressed as a Lebesgue integral. This has been proven among others by Fichtenholz and Kantorovitch (1934) \cite{french}.  \\

The inconsistency of expected utility is solved by Chichilnisky (2009) \cite{chichifear} by modifying its expression and adding a second term. This time, a continuous linear functional $\phi_2$ is introduced, and its associated measure $\mu$ is additive but not $\sigma$-additive. The new form of the ranking is then:
\begin{align*}
W(f)=& \lambda\int_\mathbb{R}f(x)\phi_1(x)dx + (1-\lambda)\langle \phi_2,f\rangle \\ W(f)=& \lambda\int_\mathbb{R}f(x)\phi_1(x)dx + (1-\lambda) \int_\mathbb{R}f(x)d\mu(x)\end{align*}
where $\lambda\in\left]0,1\right[$, $\phi_1\in L_1(\mathbb{R})$ and $\mu$ a purely finitely additive measure.  \\

It has been proven in Chichilnisky (2010) \cite{chichiswan} that this new ranking is sensitive to rare events, which means that it satisfies the appropriate axiom of ``sensitivity to rare events'' \cite{chichifear}, and therefore takes into account catastrophic events. Let us build a utility function of this kind, in order to see its evolution according to the probability $p$ of picking the deadly pill.  \\

Let us take a simple framework: the preferences are linear. The preferences function is
\begin{align*}
f\colon & \mathbb{R}\longrightarrow\mathbb{R}^+\\
f(x)=&\begin{cases}
r &\text{if $x<1$}\\
l+r &\text{if $x\geq 1$}
\end{cases}
\end{align*}
where $l$ is the unknown value of life.

The ``Lebesgue'' part of the utility function shall be:
$$ \phi_1(x)=p\times \chi _{[0,1/p]}(x)$$
Where $\chi_A$ is the characteristic function of the set $A$. The $\sigma$-additive measure associated with $\phi_1$ is the uniform probability on $[0,1/p]$, so that there is a probability $p$ to pick the $[0,1]$ interval.  \\

At this stage, the results obtained would be the same that in Section \ref{mean}, by reasoning with the mean gain. Let us add the purely finitely additive part. Let $\mu$ be the following measure: \\

\[ \forall A\subset\mathbb{R},\quad \mu(A)= \begin{cases}1& \text{if $\exists\ \mathcal{N}(o)\subset A$} \\ 0 &\text{otherwise}\end{cases} \]

$\mathcal{N}(0)$ is a ``neighbourhood'' of $0$. The first condition can be understood as ``there exists an open interval centered on 0 that is entirely included in $A$''. By using this definition, the measure $0$ is assigned to the singleton $\{0\}$, so that the measure $\mu$ is still absolutely continuous with respect to Lebesgue measure. \\

In \cite{chichiswan}, Chichilnisky shows that the measure $\mu$ is a purely finite additive measure. The explicit definition of its associated linear functional $\phi_2$ on $L_\infty(\mathbb{R})$ is only given on the subspace $CL_\infty$ of all continuous functions $f\in L_\infty(\mathbb{R})$ that have a limit towards 0.

\[
\forall f\in CL_\infty,\quad \langle\phi_2,f\rangle = \lim_{x\rightarrow 0} f(x)
\]

Since the chosen preferences are continuous at 0, the new utility function is now:
\begin{align*}
W(f)=&\lambda\int_\mathbb{R}f(x)\phi_1(x)dx + (1-\lambda)\langle \phi_2,f\rangle \\
W(f)=& \lambda \left[pr+(1-p)(l+r)\right] + (1-\lambda)\times \underbrace{\lim_{x\rightarrow 0} f(x)}_{r}
\end{align*}

\subsubsection{Application and numerical values}

The value of $W(f)$ has to be compared to $l$, in order to know if the game is worth playing or not. As a reminder:
\begin{enumerate}[label=--]
\item $l$ = \$3,000,000
\item $r$ = \$220,000
\item $p$ = $10^{-9}$
\end{enumerate}

Let us write the utility of the deal as a function of $p$ which depends on a parameter $\lambda\in]0,1[$ : 

\begin{equation}
W_\lambda (p)= \lambda \left[pr+(1-p)(l+r)\right] + (1-\lambda)r
\end{equation}

As in \eqref{limit}, let us see the evolution of $W$ when the probability of the deadly pill converges towards 0 (i.e.\ when the size of the pile expands towards infinity).

\begin{equation*}
\lim_{p\rightarrow 0}W_\lambda(p)=\lim_{p\rightarrow 0}\lambda \left[pr+(1-p)(l+r)\right] + (1-\lambda)r
\end{equation*}

\begin{equation}
\lim_{p\rightarrow 0}W_\lambda(p) = \lambda (l+r) + (1-\lambda)r = \lambda l + r
\end{equation}

Since $\lambda<1$, it is impossible to determine at first if this limit is greater or smaller than $l$: it depends on the value of $\lambda$. If $\lambda l+r<l$, the deal cannot be acceptable, regardless of how small $p$ is. If $\lambda l+r>l$, sometimes the deal will be acceptable, sometimes it will not (depending on the probability $p$).  \\

With the present parameters, the threshold is $\displaystyle\lambda_0=\frac{l-r}{l}\approx 0.926$: if $\lambda$ is smaller than this value, the subject will never accept the deal. If $\lambda$ is greater, sometimes they will, sometimes they will not.  \\

The following graph shows evolution of the value of the deal with respect to $p$, with two different given values of $\lambda$: one is greater than $\lambda_0$, the other is smaller. \\



It is important to remark that $\lambda$ is an intrinsic parameter of the utility function, while $p$ is a parameter of the deal. Changing $\lambda$ means changing the way the agent evaluates risk. Different values of $\lambda$ can explain the different behaviours of the subjects: some accept the deal, others find the probability of death too high, and others do not accept, regardless of how small $p$ can be. \\

\section{Concluding remarks}

Various hypotheses have been raised in order to explain why some people refuse some deals which have positive mean gains. In some cases, risk aversion could be a solution: there exists concave expected utility functions that explain the results obtained with some subjects during the pill experiment. Some of these functions give coherent results with smaller lotteries too. \\

Some behaviours cannot however be described using expected utility theory. By using a theory wich better takes into account catastrophic events, it is possible to explain the experimental results, including those which are inconsistent with expected utility theory. \\


\end{document}